\documentclass[prl,twocolumn,showpacs,preprintnumbers,amsmath,amssymb]{revtex4}

\usepackage{epsfig}
\usepackage{graphicx}
\usepackage{color}
\usepackage{amsmath}

\begin{document}

\title{Electrostatic extraction of cold molecules from a cryogenic reservoir}
\author{L.D. van Buuren, C. Sommer, M. Motsch, S. Pohle, M. Schenk, J. Bayerl, P.W.H. Pinkse, and G. Rempe}
\affiliation{Max-Planck-Institut f\"ur Quantenoptik,
Hans-Kopfermann-Str. 1, D-85748 Garching, Germany}

\begin{abstract}
We present a method which delivers a continuous, high-density beam of slow and internally cold polar molecules. In our source, warm molecules are first cooled by collisions with a cryogenic helium buffer gas. Cold molecules are then extracted by means of an electrostatic quadrupole guide. For ND$_3$ the source produces fluxes up to $(7 \pm ^{7} _{4}) \times 10^{10}$ molecules/s with peak densities up to $(1.0 \pm ^{1.0} _{0.6}) \times 10^9$ molecules/cm$^3$. For H$_2$CO the population of rovibrational states is monitored by depletion spectroscopy, resulting in single-state populations up to $(82 \pm 10)\%$.



\end{abstract}

\maketitle
\vspace{1cm}

Cold polar molecules open up fascinating research possibilities \cite{Dulieu2006}, such as the investigation of fundamental physics or the implementation of quantum computing protocols. Dense samples of cold molecules will allow one to study collisions and chemistry at low temperatures. This will make it possible to mimic conditions similar to those in interstellar clouds and to determine formation rates of molecules \cite{Snow2006}. With external fields, collisions and chemical rates can be manipulated, due to the Stark (Zeeman) effect of an electric (magnetic) field on the electric (magnetic) dipole moment of the molecules \cite{Krems2005}. At sub-kelvin temperatures quantum effects will dominate and resonances are expected in the collision cross section.

In general, such experiments require high-density samples of state-selected cold molecules. As the complex internal structure of molecules precludes standard laser cooling techniques, other approaches to produce cold molecules are being pursued. On the one hand, ultracold alkali dimers are created from laser-cooled alkali atoms using photo-association \cite{Jones2006} or Feshbach resonances \cite{Koehler2006}. On the other hand, methods like buffer-gas cooling \cite{Weinstein1998}, deceleration \cite{vandeMeerakker2008} and Stark filtering \cite{Junglen2004} have been developed to bring molecules directly from room temperature into the cold regime. In contrast to the indirect methods employing alkali precursor atoms, the direct methods are applicable to a much larger group of molecules.

We now report on a combination of two direct methods, cryogenic buffer-gas cooling and Stark guiding and filtering, to produce a dense and slow beam of polar molecules in predominantly one rovibrational state. The buffer gas is ideal to produce a high-density ensemble of cold molecules. The electric guide is ideal to skim cold molecules from the buffer gas and transport them to a remote location, e.g., for electric trapping. The ability to separate the cooling region from a science region avoids many complications, such as trap losses from collisions with the buffer gas and discharges in the buffer gas. Note that molecules with a magnetic dipole moment have already been extracted from a buffer-gas cell by means of permanent magnets \cite{Patterson2007}. Our method is different because it addresses the much wider class of molecules with an electric dipole moment. Moreover, electric guides are more versatile than magnetic guides. For example, they can guide low- \cite{Junglen2004} and high-field seeking \cite{Junglen2004_2,Filsinger2008} molecules and can straightforwardly be combined with electric traps \cite{Rieger2005}. The stream of cold molecules can be delivered either continuously or pulsed by controlling the voltages of the guiding electrodes. In this way the molecular beam can be adjusted to meet the demands of the experiment. In fact, an electric guide has already been used as a source for ion-molecule collision experiments \cite{Willitsch2008}. But so far all electric guides were operated with molecular sources at temperatures above 150\,K where many rovibrational states are populated. The addition of a cryogenic buffer gas strongly reduces the number of populated rovibrational states, depending on how well the molecules have thermalized with the buffer gas. In our experiment, this thermalization is monitored by a laser depletion experiment similar to the one in \cite{motsch2007}.

\begin{figure}
  \includegraphics[width=6cm]{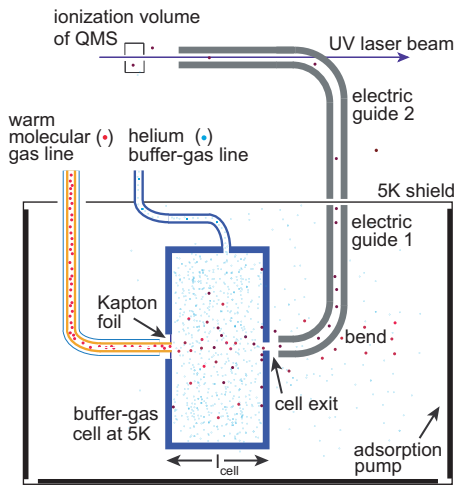}\\
  \caption{(color online). Schematic view of the experiment. Warm molecules are cooled by collisions with a cold helium gas. Behind the cell an electric quadrupole extracts slow molecules in low-field-seeking states out of the cryogenic environment into the detection chamber, where they are detected by a QMS. Depletion measurements are performed by counterpropagating an UV laser beam through the guide.
  }\label{setup}
\end{figure}

\begin{figure}
  \includegraphics[width=8.5cm]{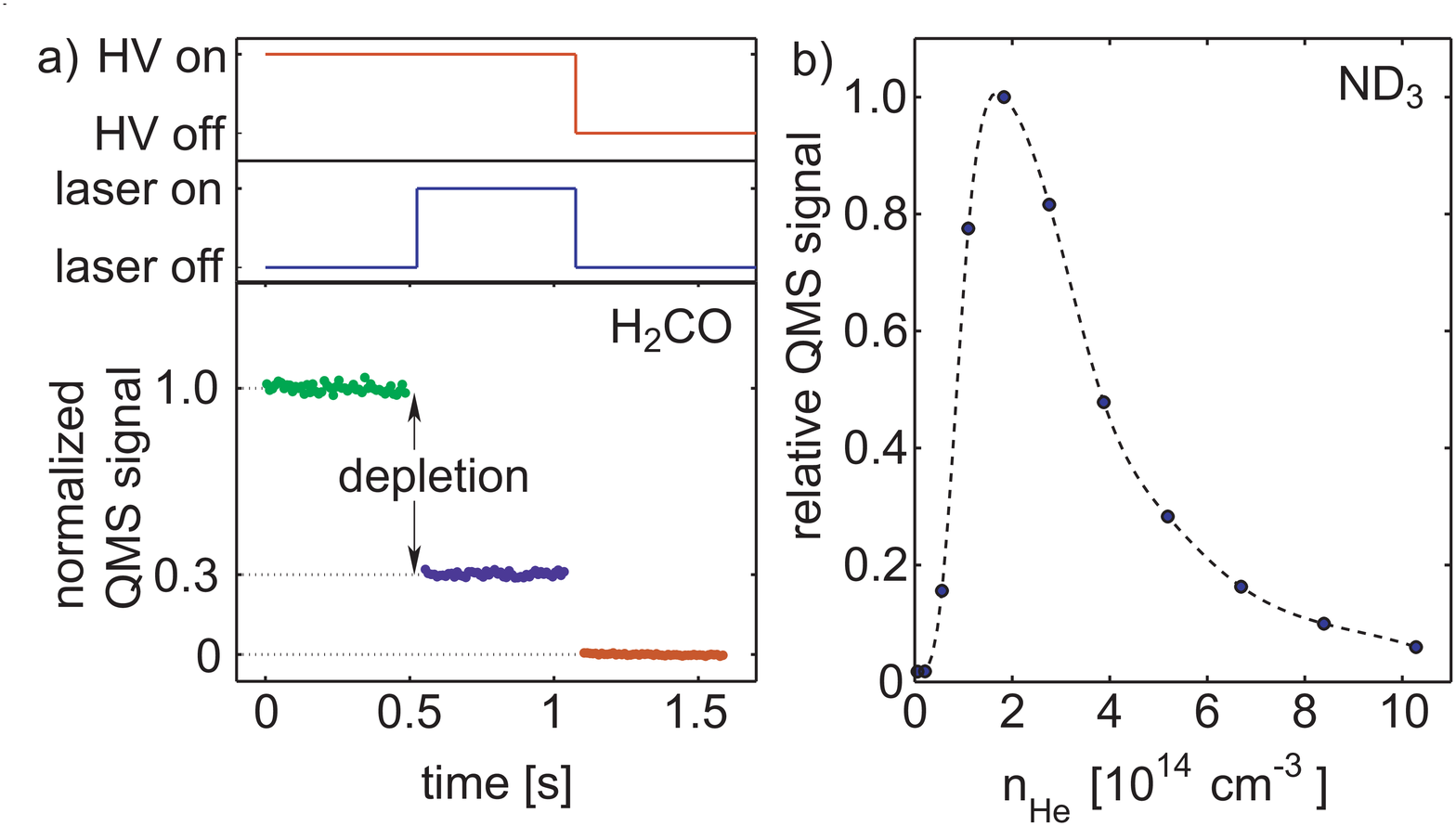}\\
  \caption{(color online). (a) Recording cycle to measure guided molecules and the depletion signal. The data are background subtracted and normalized to the average count rate without laser light.
  (b) QMS signal (normalized to the maximal value) as function of $n_{\rm He}$. For ND$_3$, the highest signal is obtained with $n_{\rm He}=1.8 \times 10^{-14}$ cm$^{-3}$. The distribution looks similar for H$_2$CO. The dashed curve is a guide to the eye.
  }\label{depl_signal}
\end{figure}

In our source, warm molecules \cite{Tnote} are passed through a cell (see Fig.\,\ref{setup}), where they are cooled by collisions with a cryogenic helium gas as cold as 5\,K.
Because of these collisions not only the velocity distribution is compressed and shifted to lower values, but also the internal (rotational) temperature of the molecules is reduced as observed in previous buffer-gas experiments \cite{Maxwell2005,Egorov2004}. We typically flow $(2-30) \times 10^{15}$ He/s and $(1-10) \times 10^{17}$\,molecules/s into the cell. Most of the molecules hit the cold cell wall where they stick. A fraction of the molecules exits the cell through a small aperture (diameter 1 mm). At 1\,mm distance, the transversely slow molecules in low-field seeking states are captured in an electric quadrupole guide similar to the one described in \cite{Junglen2004}. By bending the quadrupole guide (radius of curvature 25 mm) a selection in longitudinal velocity is made. Cryogenic temperatures are reached by using a pulse-tube refrigerator. It has 2 stages, reaching temperatures T$_1$ $\approx$ 35\,K and T$_2$ $\approx$ 5\,K, to which two gold-coated copper shields are installed to block thermal radiation. Collisions between guided molecules and background helium gas are minimized by a large area (A $\approx$ $10^3$\,cm$^2$) charcoal adsorption pump epoxied on the inner side of the second-stage radiation shield. The guide transports the molecules to a separate room-temperature vacuum system, where they are detected by a quadrupole mass spectrometer (QMS). The signal of guided molecules is the difference in steady-state count rate when switching on and off the high voltage (HV) in electric guide 1 [see Fig.\,\ref{depl_signal}(a)]. The QMS response is calibrated by monitoring the increase in count rate and pressure when injecting small flows of molecules from the gas-handling system directly into the detection region. The geometry of this cross-beam-type QMS is such that it allows for free passage of a laser beam through the straight section of the guide. Information on the purity of the guided beam is obtained from depletion measurements \cite{motsch2007}, where the molecular beam in the guide is overlapped
with an UV laser beam. When tuned to a molecular transition, the measured decrease in count rate reflects the population of the relevant molecular state as shown in Fig.\,\ref{depl_signal}(a).

The cooling and guiding technique presented here works for all polar species which possess reasonably large Stark shifts, and which can be introduced into the buffer-gas cell. We have used deuterated ammonia (ND$_3$) to optimize the flux from the source. In these measurements, we have varied the gas flows of both the helium and the molecular gas as well as the cell length ($l_{\rm cell}$ = 1, 2, and 3\,cm) at guide voltages of $\pm$ 5\,kV, corresponding to an electric field of 93\,kV/cm. The highest flux is expected with a high density of helium gas ($n_{\rm He}$) in a short cell, such that the mean free path $\lambda$ inside the buffer-gas cell is of the order of the diameter of the exit aperture ($\varnothing =1$\,mm) \cite{Maxwell2005}. For $\lambda \ll \varnothing$ the molecules are accelerated by collisions with helium in close proximity of the exit aperture causing a smaller fraction to remain guided through the bend in the quadrupole, while for $\lambda \gg \varnothing$ only a very long cell will allow for enough collisions to slow down the molecules. But in that case most molecules will diffuse to the cell walls and stick there. We found best results for $l_{\rm cell}=2\,$cm and $n_{\rm He}$ of order $10^{14}\,$cm$^{-3}$ at which the guided ND$_3$ molecules have a measured average velocity of $\langle v \rangle$ $\approx65\,$m/s. A typical scan of the helium density in the buffer-gas cell is shown in Fig.\,\ref{depl_signal}(b). The density in the cell is estimated from the pressure in the gas line at room temperature and its measured conductance. The absolute value suffers from large systematic uncertainties, due to the indirect method used to obtain this gas line conductance. For optimal settings, the highest count rate measured by the QMS corresponds to a flux of $(7 \pm ^{7} _{4}) \times 10^{10}\,$molecules/s in the guide, which includes corrections for the limited spatial acceptance of the QMS. This flux is comparable to the one obtained previously \cite{Junglen2004}. It results in a peak density in the guide up to $(1.0 \pm ^{1.0} _{0.6}) \times 10^9\,$molecules/cm$^3$ assuming a Gaussian beam profile with $\sigma=400$\,$\mu$m \cite{Junglen2006}, where $\sigma$ represents the square root of the variance.

A feature of our setup is that we can gradually adjust the cooling of the internal degrees of freedom. The amount of cooling depends on both the temperature of the buffer-gas cell ($T_{\rm cell}$) and $n_{\rm He}$. This allows not only to study the cooling process, but also to tune the source to other preferred internal temperatures. The lowest guidable state of ND$_3$ is expected to have a population of 96\%, if we assume complete thermalization of the molecules to 5\,K. However, it is {\sl a priori} unclear how much the internal degrees have cooled inside the buffer-gas cell, if the source is optimized for maximum flux. The questions concerning state population and purity of guided molecules can be answered by laser depletion measurements as described for a similar setup in \cite{motsch2007}. Since ND$_3$ has no easily accessible optical transition, we performed these measurements with formaldehyde (H$_2$CO). Switching these gases required only minor changes in the gas handling system. The required light at wavelength 330\,nm is produced by second-harmonic generation of light from a ring dye laser. Upon excitation the molecules dissociate and are lost from the guide as shown in Fig.\,\ref{depl_signal}(a).
For these measurements the guide is switched to a voltage of $\pm 3\,$kV, which compromises between high flux and small line broadening due to the inhomogeneous electric field in the guide. The data shown here are taken at high laser power ($P_{\rm laser} \approx 100\,$mW) for which the depletion values reach $\sim$99\% of the high-power limit, determined by extrapolating the measured power dependence.

\begin{figure}
  \includegraphics[width=6.5cm]{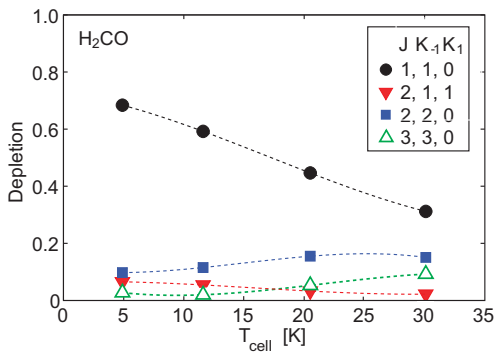}\\
  \caption{(color online). Depletion as function of $T_{\rm cell}$ for different states. The dashed lines are guides to the eye.
  }\label{depl_vs_T}
\end{figure}

The states of H$_2$CO are labeled by $|J, K_{-1}, K_1 \rangle$ in which $J$ is the rotational quantum number and $K_{-1}$ ($K_1$) its projection along the molecular axis for the prolate (oblate) limiting case \cite{Townes1975}. Because of varying electric field strength and orientation along the laser beam, all guidable substates with different $M$-values, where $M$ is the projection of the angular momentum on the electric field direction, contribute to the measured depletion signal. We have simulated the experiments assuming molecules are extracted from a thermal ensemble at temperature $T$. Nuclear spin statistics and Stark shifts of the rotational states in the guiding fields are included. The Stark shifts are calculated by numerical diagonalization of the asymmetric rotor Hamiltonian in the presence of an electric field \cite{motsch2007}. During the cooling in the inert buffer gas the ortho and para states of H$_2$CO do not mix. From simulations at $T = 5\,$K, we expect contributions higher than 1\% from the $|1,0,1\rangle$, $|1,1,0\rangle$, and $|2,1,1 \rangle$ low-field seeking states. These are measured together with the $|2,2,0\rangle$ and $|3,3,0\rangle$ states, which possess relatively large Stark shifts and which contribute at higher temperatures. The depletion signal changes for all measured states as function of $T_{\rm cell}$ as shown in Fig.\,\ref{depl_vs_T}.
Highest populations are found for the $|1,1,0\rangle$ state. At $T_{\rm cell} \approx 5$ K no contribution of the  $|1,0,1 \rangle$ state is found. Since this state is expected to have its highest contribution at our minimal $T_{\rm cell}$, due to its low rotational energy of 2.4\,cm$^{-1}$, we did not probe it in the other measurements. For the data shown in Fig.\,\ref{depl_vs_T}, the pressure in the buffer-gas line was kept constant at room temperature at the optimal settings found in Fig.\,\ref{depl_signal}, causing variations in $n_{\rm He}$ depending on $T_{\rm cell}$. Still, the increase of the $|2,2,0 \rangle$ and $|3,3,0 \rangle$ data (rotational energies of 40\,cm$^{-1}$ and 88\,cm$^{-1}$, resp.) and the decrease of the $|1,1,0 \rangle$ and $|2,1,1 \rangle$ data (rotational energies of 11\,cm$^{-1}$ and 16\,cm$^{-1}$, resp.) indicate clearly a dependence of the rotational temperature on the temperature of the buffer gas. From the data we cannot conclude whether the molecules completely thermalize inside the buffer-gas cell. Therefore, an increase in the $|1,1,0 \rangle$ population might be feasible by increasing $n_{\rm He}$ at minimal $T_{\rm cell}$, as discussed below.

\begin{figure}
  \includegraphics[width=6.5cm]{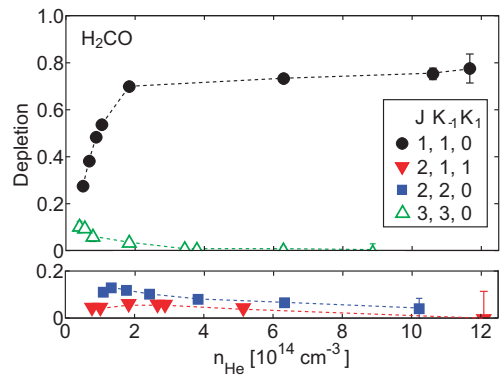}\\
  \caption{(color online). Depletion as function of $n_{\rm He}$ for different states. The large statistical errorbars at high $n_{\rm He}$ are caused by the relatively low count rates at these settings. The lines are guides to the eye.
  }\label{depl_vs_nHe}
\end{figure}

Figure\,\ref{depl_vs_nHe} presents depletion measurements as function of $n_{\rm He}$ at $T_{\rm cell} \approx 5$ K. Again, highest populations are found for the $|1,1,0\rangle$ state. At low densities, the population increases strongly with $n_{\rm He}$, caused by the onset of rotational cooling of the molecules by collisions with cold helium gas. Also a decrease in the relatively high-energetic $|3,3,0\rangle$ state is visible. For $n_{\rm He} \gtrsim 3 \times 10^{14}\,$cm$^{-3}$ the populations converge to constant values. In this region, the total depletion signal sums up to 80-90\%, illustrating the low number of states in the guided beam. The remaining 10-20\% is attributed to molecular orbits which do not interact with the laser beam and to molecules which do not dissociate, but decay back into guidable states. A similar fraction was found in our previous experiments performed at room temperature \cite{motsch2007}. Taking this into account the population in the $|1,1,0\rangle$ state of H$_2$CO is (82$\pm$10)\% for the optimal flux setting of $n_{\rm He}$.

A possible explanation for the absence of the $|1,0,1\rangle$ state in the beam is given by its relatively low Stark shift. We know from velocity distributions \cite{Vnote} at high $n_{\rm He}$, that the molecules can be accelerated when leaving the cell (this is called velocity boosting). Therefore, this Stark shift might be too low to keep the molecules guided in the quadrupole bend. At low $n_{\rm He}$ the velocity boost is small, but in this case data ($n_{\rm He} \lesssim 2 \times 10^{14}\,$cm$^{-3}$ in Fig.\,\ref{depl_vs_nHe}) indicate higher rotational temperatures for which the population in $|1,0,1\rangle$ is decreasing rapidly.

In general, the addition of buffer-gas cooling to the electric filtering method purifies the extracted molecular beam. This is especially beneficial for molecules with linear Stark shifts, because these molecules populate many guidable states at room temperature. For low-mass molecules with quadratic Stark shifts, such as water, the number of states contributing to an electrically filtered beam is much less than for the linear case \cite{Rieger2006,Motsch2008}. Still, our source can enhance the purity of the extracted beam by adjusting the temperature of the buffer-gas cell to the rotational temperature of a low-lying state with large Stark shift. Currently, the source is being employed to produce beams consisting of polar molecules with masses up to $\sim$ 100 amu, showing the generality of our method.

In summary, we have developed a source which extracts a continuous beam of slow molecules out of a cryogenic buffer gas and delivers them into a separate ultrahigh vacuum chamber. The source can operate for many hours with peak densities inside the quadrupole guide up to $(1.0 \pm ^{1.0} _{0.6}) \times 10^9\,$molecules/cm$^3$ for ND$_3$. The internal temperature of the molecules is reduced by collisions with the helium gas. This leads to a strong increase of the purity of the beam.

The extracted beam can be employed to load electric traps \cite{Rieger2005} in which collisions and chemistry at low temperature can be observed. Such experiments might also be performed by crossing separate beams of cold molecules. Besides, collisions between cold molecules and alkali atoms stored in an AC electric trap \cite{Schlunk2007,Rieger2007} can be studied. Because of the high purity of our beam, these processes can be investigated in a state-selective manner.
Our source can also be employed to improve the resolution in high-precision experiments \cite{EDM}.

In future experiments, the flux might be enhanced by pumping the (high-field seeking) ground-state population to a selected low-field seeking state in the region between the buffer-gas cell and the guide \cite{vandeMeerakker2003}.
The number of slow molecules itself can be increased by lowering the temperature of the buffer gas.
This will further increase the purity of the beam in particular for heavier molecules with smaller
rotational constants.
If required, higher purities can also be obtained by laser depletion of unwanted states.
Finally, extraction of high-field seeking states could be achieved by AC voltages on the guiding electrodes \cite{Junglen2004_2,Filsinger2008}. This would, for example, enable the production of a high flux, guided beam of slow ground-state molecules.

\begin{acknowledgments}
Support by the Deutsche Forschungsgemeinschaft through the excellence cluster "Munich Centre for Advanced Photonics" and EuroQUAM (Cavity-Mediated Molecular Cooling) is acknowledged.
\end{acknowledgments}



\end{document}